\documentclass[fleqn,12pt,twoside,a4paper]{article}
\usepackage[headings]{espcrc1}

\usepackage{graphicx}
\usepackage{amssymb}

\newcommand{\qs}{Q_{\mathrm{s}}}
\newcommand{\ra}{R_A}
\newcommand{\nf}{N_\mathrm{f}}

\newcommand{\ud}{\mathrm{d}}
\newcommand{\xt}{\mathbf{x}_T}
\newcommand{\yt}{\mathbf{y}_T}

\newcommand{\qt}{{\mathbf{q}_T}}

\newcommand{\gev}{\textrm{ GeV}}
\newcommand{\fm}{\textrm{ fm}}

\title{Quark-antiquark production from classical fields and chemical equilibration}

\author{F. Gelis,\address{Service de Physique Th\'eorique, 
B\^at. 774, CEA/DSM/Saclay, 91191 Gif-sur-Yvette, France}
K. Kajantie\address[HY]{Department of Physics, P.O.Box 64,
FI-00014 University of Helsinki, Finland}
and
T. Lappi\addressmark[HY]\address{Helsinki
Institute of Physics, P.O.Box 64, FI-00014 University of Helsinki,
Finland}
}

\begin{document}

\maketitle

\begin{abstract}
We compute by numerical integration of the Dirac equation the number of
quark-antiquark pairs produced in the classical color fields of colliding
ultrarelativistic nuclei. The backreaction of the created pairs on the color fields is
not taken into account. While the number of $q\bar q$ pairs is parametrically
suppressed in the coupling constant, we find that in this classical field model it
could even be compatible with the thermal ratio to the number of gluons. After
isotropization one could thus have quark-gluon plasma in chemical equilibrium.
\end{abstract}

\section{Introduction}

The initial stages of an ultrarelativistic heavy ion collision are believed to
be dominated by strong classical color fields.
There is a twofold interest in calculating the production of quark--qntiquark pairs
from these classical fields. Firstly, although heavy quark production is 
in principle calculable perturbatively, it would be interesting to understand
whether these strong color fields influence the result. Secondly, being able 
to compute both gluon and quark production in the same framework would give
insight into the chemical equilibration of the system and test the consistency
of the assumption of gluon dominance. 
The number of quark pairs present in the early stages of the system
has observable consequences in the thermal photon and dilepton spectrum.

In this talk we shall present first results \cite{Gelis:2005pb} of a numerical computation 
of quark antiquark pair production from the classical fields of the 
McLerran-Venugopalan (MV) model. The equivalent calculation, although in another
gauge, has been carried out analytically to lowest order in the densities of both 
color sources (``pp''-case)  in Ref.~\cite{Gelis:2003vh} and to lowest 
order in one of the sources (``pA''-case) in Ref.~\cite{Fujii:2005vj}.
The corresponding calculation in the Abelian theory \cite{Baltz:1998zb,Baltz:2001dp}, 
of interest for the physics of ultraperipheral collisions,
can be done analytically to all orders in the electrical charge of the
nuclei. Quark pair production has also been studied in a related
``CGC''-approach in Refs.~\cite{Kharzeev:2003sk,Tuchin:2004rb} and in a more
general background field in Ref.~\cite{Dietrich:2004eb}.

\section{The numerical calculation}

Our calculation of pair production relies on the numerical calculation of 
the classical background color field in which
we solve the Dirac equation.

In the classical field model the background gluon field is obtained from solving
the Yang-Mills equation of motion with the classical color source $J^{\nu}$ given by transverse
color charge distributions of the two nuclei boosted to infinite energy:
\begin{equation}
[D_{\mu},F^{\mu \nu}] = J^{\nu}
=\delta^{\nu +}\rho_{(1)}(\xt)\delta(x^-)
+ \delta^{\nu -}\rho_{(2)}(\xt)\delta(x^+).
\end{equation}
In the MV model \cite{McLerran:1994ni} the color charges are taken as
random variables with a Gaussian distribution
\begin{equation}
\langle \rho^a(\xt) \rho^b(\yt) \rangle 
= g^2 \mu^2 \delta^{ab}\delta^2(\xt-\yt),
\end{equation}
depending on the coupling $g$  and a phenomenological parameter $\mu$. The combination
$g^2\mu$ is closely related to the saturation scale
$\qs$. Collisions of two ions were first studied analytically using this model
in  Ref.~\cite{Kovner:1995ts} and the way of numerically solving the 
equations of motion was formulated in Ref.~\cite{Krasnitz:1998ns}.

\begin{figure}
\begin{minipage}[t]{0.44\textwidth}
\begin{center}
\includegraphics[height=6cm]{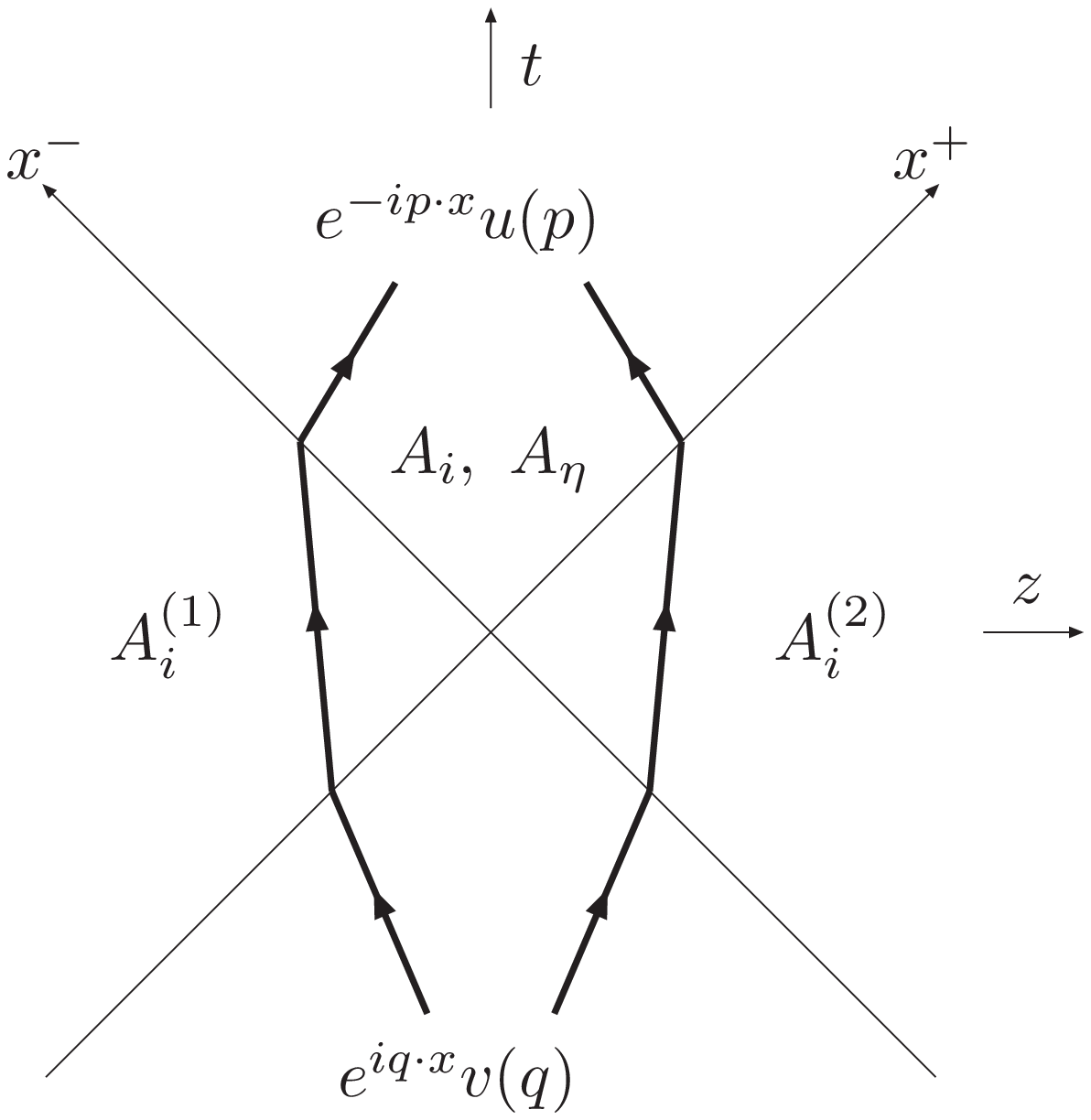}
\end{center}
\vspace{-1cm}
\caption{
Domains of different time dependences. The fermion amplitude
is a sum of two terms: one with interaction first with the left moving
nucleus, then the right moving one, and vice versa.
$A_i^{(1,2)}$ are pure gauges
and $A_i,A_\eta$ is a numerically computed color field.
}\label{fig:spacet}
\end{minipage}
\hfill
\begin{minipage}[t]{0.54\textwidth}
\begin{center}
\includegraphics[height=5.5cm]{taudep2gevplusbp.eps}
\rule{0cm}{0.5cm}
\end{center}
\vspace{-1cm}
\caption{
Dependence on proper time $\tau$ of the number of pairs
of one flavor per unit rapidity $\ud N/\ud y$
for $g^2\mu=2$ GeV and for values of quark mass marked on the
figure. The lowest curve corresponds to $g^2\mu=1$ GeV.
}
\label{fig:taudep}
\end{minipage}
\end{figure}

Our method of solving the Dirac equation is explained in more detail 
and the numerics tested in a 1+1-dimensional toy  model
in Ref.~\cite{Gelis:2004jp}. The domains of spacetime involved in the calculation
are illustrated in Fig.~\ref{fig:spacet}.
One starts in the the infinite past $t\to -\infty$ with a negative
energy plane wave solution $\psi(x) = e^{i q \cdot x}v(q)$. The Dirac equation
can then be integrated forward in time analytically to the future light cone
($\tau^2 = 2 x^+x^-=0, \ x^\pm > 0$) because the background field in the intermediate region
is a pure gauge. This gives an initial condition for numerically solving the Dirac 
equation for $\tau \ge 0$ using the coordinate system  $\tau,z,\xt$. The resulting
spinor wavefunction $\psi(\tau,z,\xt)$ is then projected onto positive
energy states $e^{-ip\cdot x}u(p)$ at time $\tau$ to obtain the amplitude $M_\tau$.
For times larger than the formation time of the quark pair 
$\tau \gtrsim 1/\sqrt{m^2 + \qt^2}$ this amplitude can be interpreted as the amplitude 
for producing quark antiquark pairs. The resulting number of quark pairs is shown 
in Fig.~\ref{fig:taudep}.

\begin{figure}
\begin{minipage}[t]{0.48\textwidth}
\includegraphics[width=0.99\textwidth]{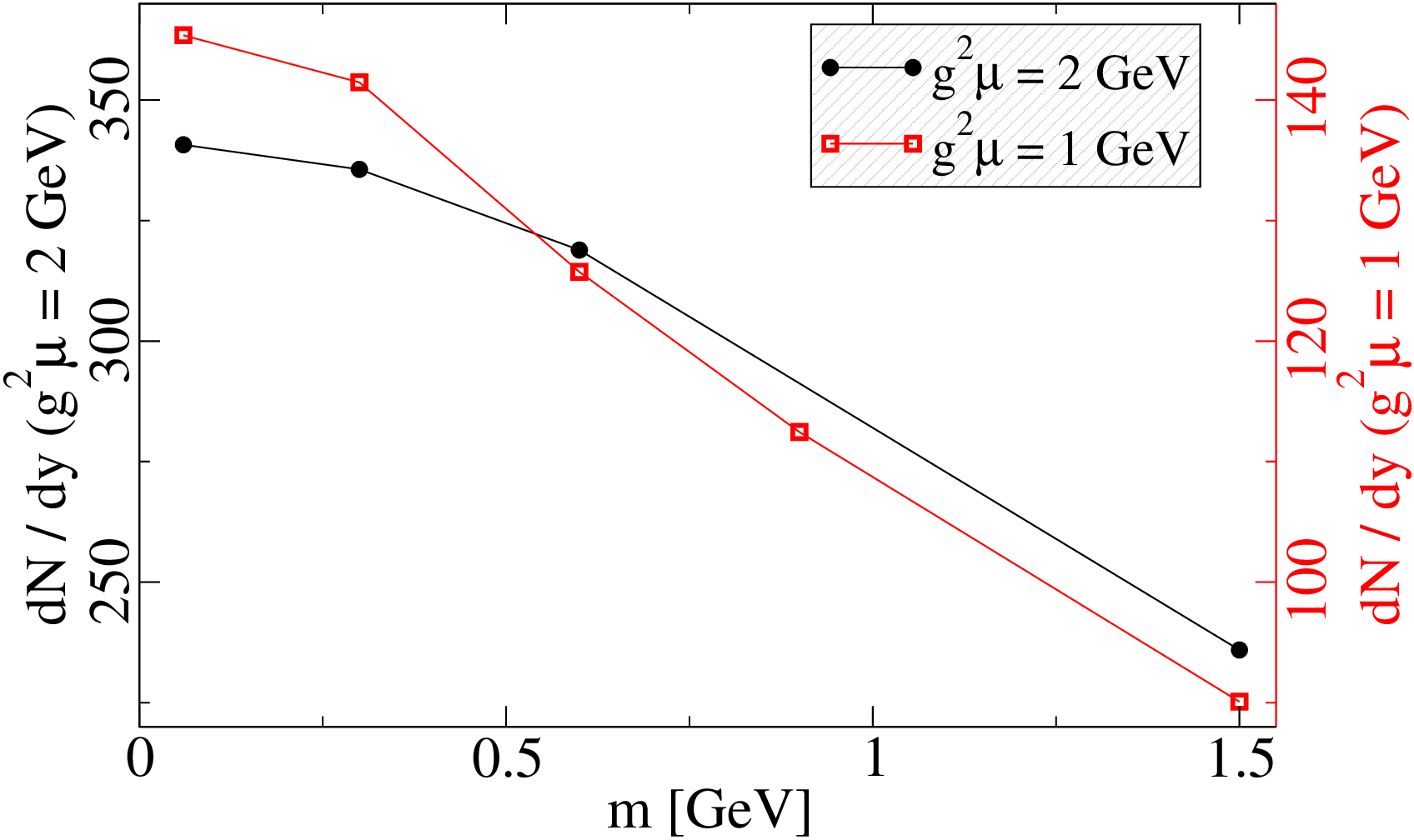}
\vspace{-1cm}
\caption{
Dependence of the number of quark pairs
on quark mass at a fixed proper time, $\tau=0.25 \textrm{ fm}$,
and for two values of $g^2\mu$.
}
\label{fig:gsqrmu}
\end{minipage}
\hfill
\begin{minipage}[t]{0.48\textwidth}
\includegraphics[width=0.99\textwidth]{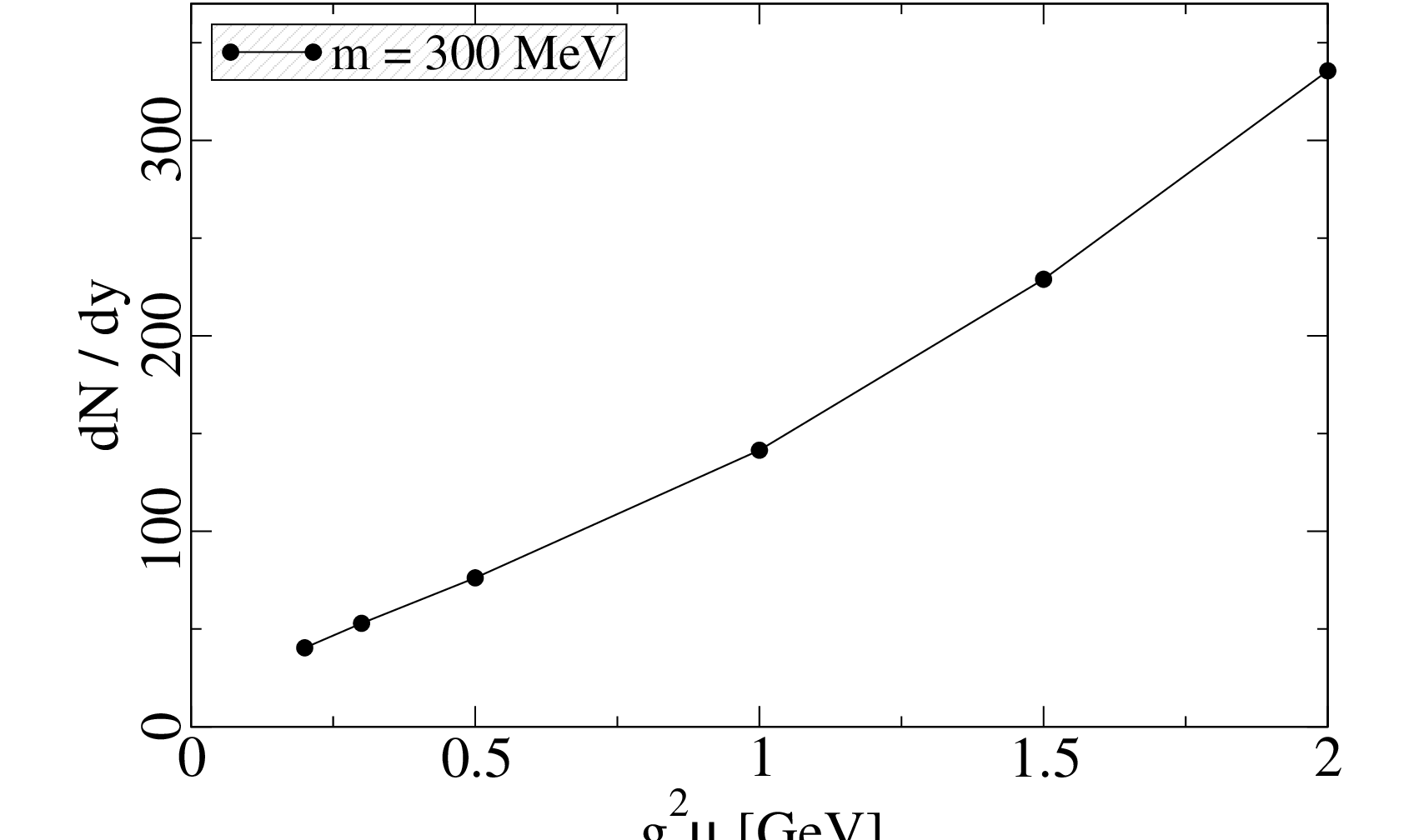}
\vspace{-1cm}
\caption{
Dependence of the number of quark pairs on $g^2\mu$ at a
fixed proper time, $\tau=0.25 \textrm{ fm}$, and for quark mass $m=0.3\textrm{ GeV}$.
} 
\label{fig:mass}
\end{minipage}
\vspace{-1cm}
\end{figure}

\begin{figure}
\begin{minipage}[t]{0.48\textwidth}
\includegraphics[width=0.99\textwidth]{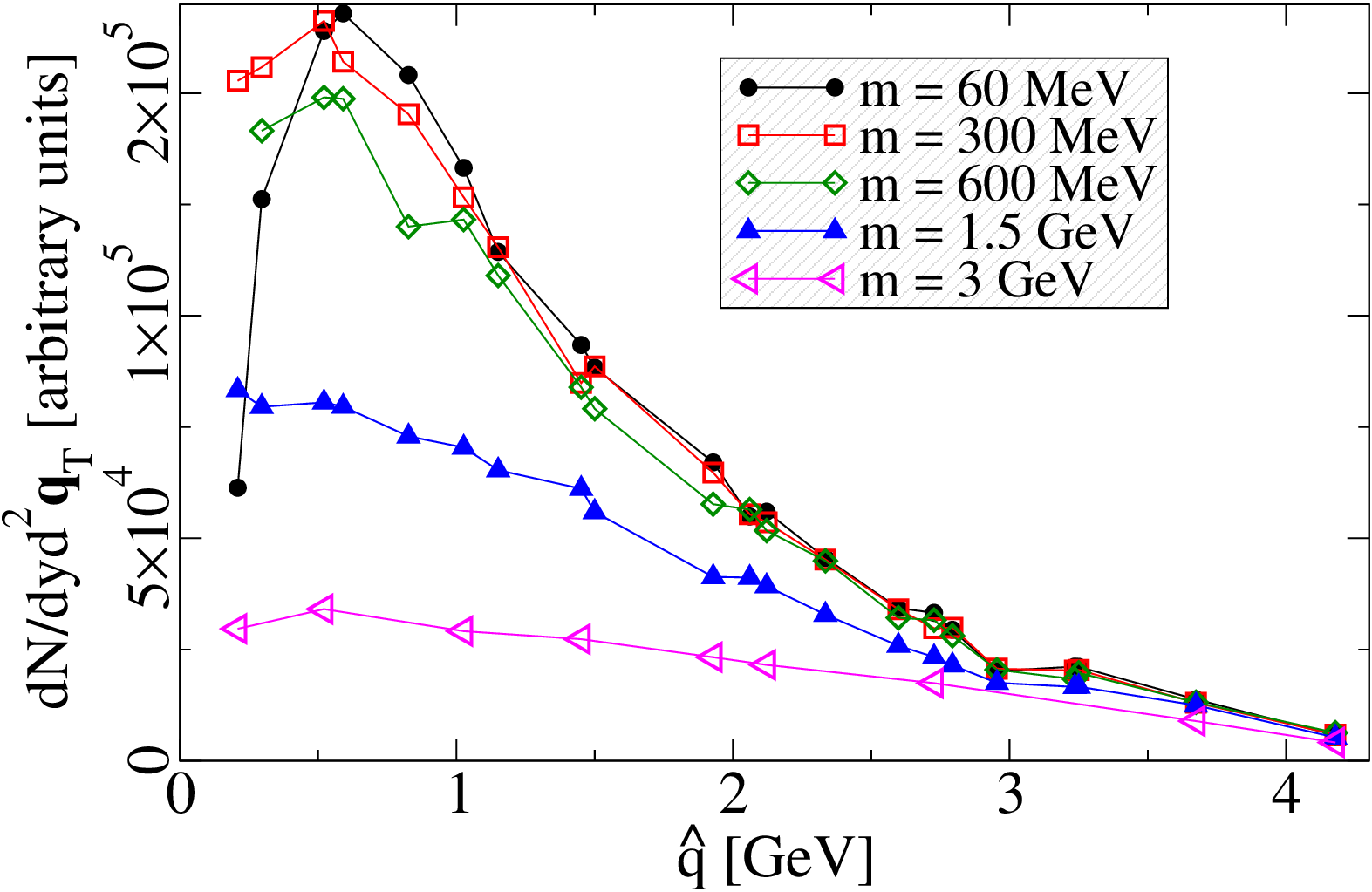}
\vspace{-1cm}
\caption{
Transverse momentum spectrum of (anti)quarks for
$g^2\mu=2$ GeV at a fixed proper time, $\tau=0.25$ fm, and
for different quark masses.
}
\label{fig:spectsm}
\end{minipage}
\hfill
\begin{minipage}[t]{0.48\textwidth}
\includegraphics[width=0.99\textwidth]{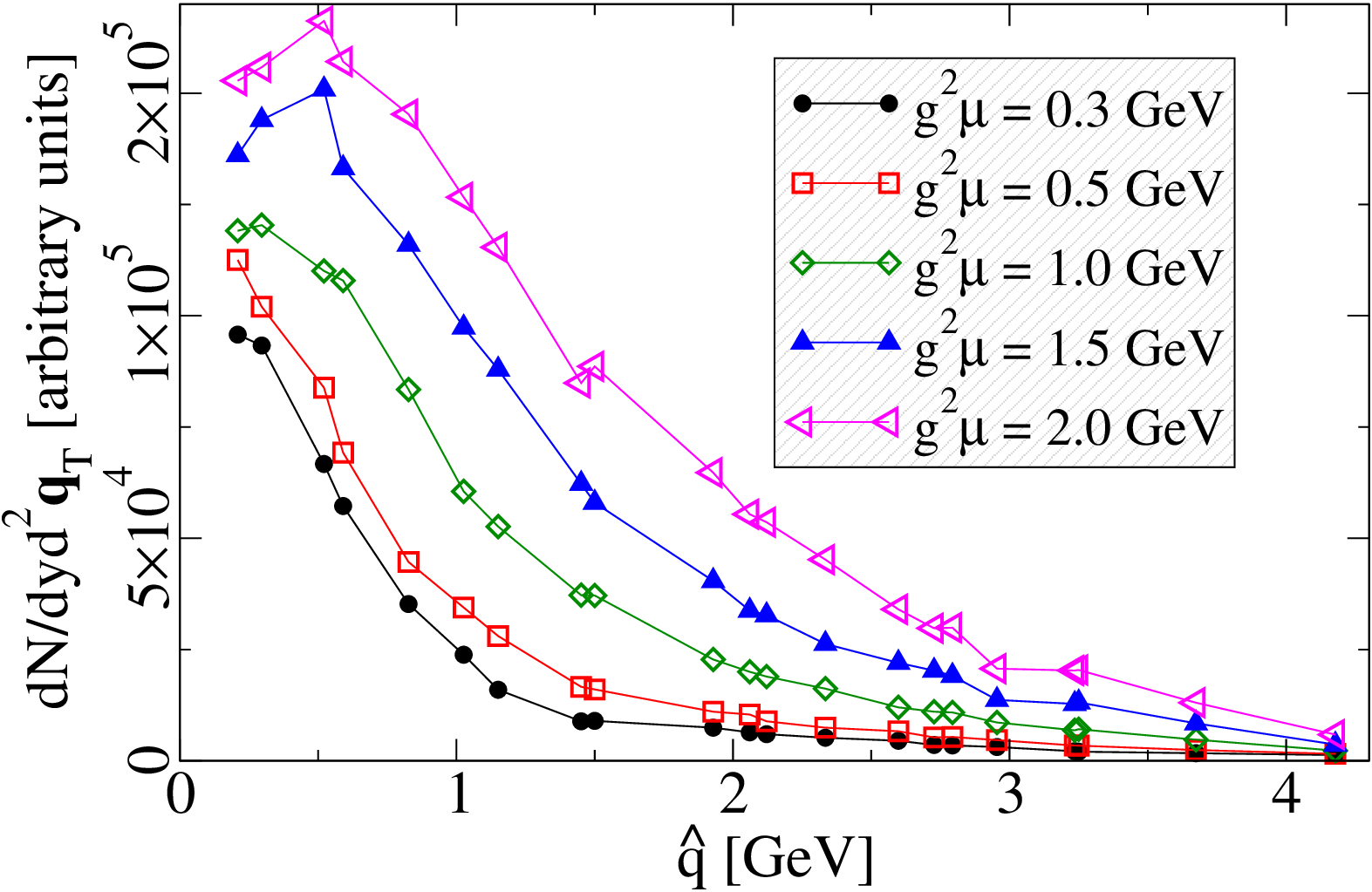}
\vspace{-1cm}
\caption{
 Transverse momentum spectrum of (anti)quarks for quark mass
$m=0.3$ GeV and for different $g^2\mu$ at a fixed proper time, $\tau=0.25$ fm.
} 
\label{fig:spectsgsqrmu}
\end{minipage}
\end{figure}

The physical parameters of the calculation are  $g^2\mu$ characterising the
strength of the background field, the nuclear radius $\ra$ and  the quark mass $m$.
The dependence on $g^2\mu$ and $m$ of the number of pairs at $\tau=0.25 \fm$ is shown in
Figs.~\ref{fig:gsqrmu} and \ref{fig:mass}. 
The transverse momentum spectra of the (anti)quarks as a 
function of $\qt$ is shown for different quark masses and saturation scales in 
Figs.~\ref{fig:spectsm}  and \ref{fig:spectsgsqrmu}. 

\section{Discussion}

According to conventional wisdom the initial state of a heavy ion collision
is dominated by gluons. Assuming that the subsequent evolution of the system 
conserves entropy this would mean $\sim 1000$  gluons in a unit of rapidity.
In the classical field model this corresponds \cite{Lappi:2003bi} to 
$g^2\mu \approx 2 \gev$. Our results seems to point to a rather large number 
of quark pairs present already in the initial state. One could envisage a scenario
where for $g^2\mu \approx 1.3 \gev$ these 1000 particles could consist of
$\gtrsim 400$ gluons, $\gtrsim 300$ quarks and $\gtrsim 300$ antiquarks
(take the lowest curve from Fig.~\ref{fig:taudep} and multiply by $\nf=3$).
This would be close to the thermal ratio of $N_g/N_q = 64/(21 \nf)$.

\section{Conclusions}

We have calculated 
quark pair production from classical background field of McLerran-Venugopalan 
model by solving the 3+1--dimensional Dirac equation numerically 
in this classical background field. We find that 
number of quarks produced is large, pointing to a possible fast
chemical equilibration of the system.
The mass dependence of our result is surprisingly weak
and we are not yet able to make any conclusions on heavy quarks until
studying the numerical issues involved.
\subsection*{Acknowledgements}
T.L. was supported by the Finnish Cultural
Foundation.  This research has also been supported by the Academy of
Finland, contract 77744. We thank  R. Venugopalan,
H. Fujii, K. J. Eskola, B. Mueller and D. Kharzeev for discussions.

\bibliographystyle{h-physrev4}
\bibliography{spires}

\begin{thebibliography}{10}

\bibitem{Gelis:2005pb}
F.~Gelis, K.~Kajantie and T.~Lappi,
\newblock hep-ph/0508229.

\bibitem{Gelis:2003vh}
F.~Gelis and R.~Venugopalan,
\newblock Phys. Rev. {\bf D69}, 014019 (2004), [hep-ph/0310090].

\bibitem{Fujii:2005vj}
H.~Fujii, F.~Gelis and R.~Venugopalan,
\newblock hep-ph/0504047.

\bibitem{Baltz:1998zb}
A.~J. Baltz and L.~D. McLerran,
\newblock Phys. Rev. {\bf C58}, 1679 (1998), [nucl-th/9804042].

\bibitem{Baltz:2001dp}
A.~J. Baltz, F.~Gelis, L.~D. McLerran and A.~Peshier,
\newblock Nucl. Phys. {\bf A695}, 395 (2001), [nucl-th/0101024].

\bibitem{Kharzeev:2003sk}
D.~Kharzeev and K.~Tuchin,
\newblock Nucl. Phys. {\bf A735}, 248 (2004), [hep-ph/0310358].

\bibitem{Tuchin:2004rb}
K.~Tuchin,
\newblock Phys. Lett. {\bf B593}, 66 (2004), [hep-ph/0401022].

\bibitem{Dietrich:2004eb}
D.~D. Dietrich,
\newblock Phys. Rev. {\bf D70}, 105009 (2004), [hep-th/0402026].

\bibitem{McLerran:1994ni}
L.~D. McLerran and R.~Venugopalan,
\newblock Phys. Rev. {\bf D49}, 2233 (1994), [hep-ph/9309289].

\bibitem{Kovner:1995ts}
A.~Kovner, L.~D. McLerran and H.~Weigert,
\newblock Phys. Rev. {\bf D52}, 3809 (1995), [hep-ph/9505320].

\bibitem{Krasnitz:1998ns}
A.~Krasnitz and R.~Venugopalan,
\newblock Nucl. Phys. {\bf B557}, 237 (1999), [hep-ph/9809433].

\bibitem{Gelis:2004jp}
F.~Gelis, K.~Kajantie and T.~Lappi,
\newblock Phys. Rev. {\bf C71}, 024904 (2005), [hep-ph/0409058].

\bibitem{Eskola:1999fc}
K.~J. Eskola, K.~Kajantie, P.~V. Ruuskanen and K.~Tuominen,
\newblock Nucl. Phys. {\bf B570}, 379 (2000), [hep-ph/9909456].

\bibitem{Lappi:2003bi}
T.~Lappi,
\newblock Phys. Rev. {\bf C67}, 054903 (2003), [hep-ph/0303076].

\end{thebibliography}

\end{document}